\documentclass[preprint,showpacs,aps,prd,nofootinbib,floatfix,amsmath,amssymb]{revtex4}
\usepackage{graphicx}
\usepackage[utf8x]{inputenc}
\usepackage{color}
\usepackage{subfig}
\usepackage{multirow}
\usepackage{slashed}

\begin{document}


\title{Radiative neutrino masses and exotic right-handed neutrinos}


\author{A. C. B. Machado$^{(a)}$} \email{ana.machado@ufabc.edu.br}
\author{J. Monta\~no$^{(b)}$} \email{ jmontano@conacyt.mx}
\author{V. Pleitez$^{(c)}$} \email{v.pleitez@unesp.br}
\author{M. C. Rodriguez$^{(d)}$} \email{marcoscrodriguez@ufrrj.br}

\affiliation{
$^{(a)}$Centro de Ci\^encias Naturais e Humanas, Universidade Federal do ABC, Santo Andr\'e - SP, 09210-170,
Brazil.
\\
$^{(b)}$C\'atedras Conacyt--Facultad de Ciencias F\'isico Matem\'aticas, Universidad Michoacana de San Nicol\'as de Hidalgo,
Av. Francisco J. M\'ugica s/n, C.~P. 58060, Morelia, Michoac\'an, M\'exico.
\\
$^{(c)}$Instituto  de F\'isica Te\'orica, Universidade Estadual Paulista, R. Dr. Bento Teobaldo Ferraz 271, Barra Funda, S\~ao Paulo - SP, 01140-070, Brazil.
\\
$^{(d)}$ Departamento de F\'\i sica,
Universidade Federal Rural do Rio de Janeiro - UFRRJ,
BR 465 Km 7, 23890-000,
Serop\'edica, RJ, Brazil
}

\date{Oct/06/2020}


\begin{abstract}

We consider an extension of the standard electroweak model with three Higgs doublets and global $B-L$ and $\mathbb{Z}_2$ symmetries. Two of the scalar doublets are inert due to the $\mathbb{Z}_2$ symmetry. We calculated all the mass spectra in the scalar and lepton sectors and accommodate the leptonic mixing matrix as well. We also include an  analysis of the scalar sector, showing that the potential is limited from below, and we obtain the masses of the scalar sector. Furthermore we consider the effects of the model on the anaomalous magnetic dipole of charged leptons and the $\mu\to e\gamma$ decay. We also present the SUSY version of the model with global $B-L$.

\end{abstract}


\maketitle

\section{Introduction}

Today we know that neutrinos have mass, that  their flavors are mixed and also that there is at least one Higgs boson~\cite{Olive:2016xmw}. However, we do not know yet if neutrinos are Majorana or Dirac particles and if there are more neutral scalars. Moreover, right-handed sterile neutrinos i.e., singlest under the standard model (SM) symmetries, have not been observed yet.

If there exist more scalars fields an interesting possibility is that only one of them contributes to the spontaneous symmetry breaking of the SM and the others are of the inert type~\cite{Ma:2015eoa}.
The existence of inert scalar fields comes from very long ago, in particular the inert doublet was first considered in Ref.~\cite{Deshpande:1977rw}. These sort of scalars are not only intresting \textit{per se}, as they can be candidates to dark matter~\cite{Fortes:2014dca}, but also because neutrino masses may arise at the 1-loop level~\cite{Ma:2006km} and, at the same time, the Pontecorvo-Maki-Nakagawa-Sakata (PMNS) matrix is also accommodated.
Here we consider a model with $SU(2)\otimes U(1)\otimes U(1)_{B-L}$ gauge symmetries but with the particle spectra enlarged with two inert doublets with the same quantum number as the standard model Higgs doublet $Y=+1$. We will denote the scalars as follows: $H\equiv S,D_1,D_2$, being the latter two doublets inert because of an unbroken $Z_2$ symmetry. We also include three right-handed neutrinos with non-standard assignment of global $B-L$ charges that make the $U(1)_{B-L}$ symmetry anomaly free~\cite{Montero:2007cd,Ma:2014qra}.

The outline of this paper is as follows: In Sec. \ref{sec:model} we present the model, in Sec. \ref{sec:leptons} we show that the model can ajust the PMNS mixing matrix and the leptons masses. The phenomenology of the model is discussed in Sec.~\ref{sec:pheno}.
In Sec.~\ref{sec:susymodel} we present the supersymmetric version of the model. Finally, our conclusions appear in the last section.

\section{The Model}
\label{sec:model}

The representation content of the model is the following: under $SU(2)_L$ we have the lepton doublets $L_i=(\nu_i l_i)^T_L,\, i=1,2,3$ and charged singlets $l_{iR}$; three sterile neutrinos $N_{iR}$; the SM Higgs $S$, and two scalar doublets $D_{1,2}$ all of them with $Y=+1$.
The Yukawa interactions are given by
\begin{eqnarray}
\label{yukawa}
-\mathcal{L}^{leptons}_{Yukawa}&=& G^l_{ij}\bar{L}_{iL}l_{jR}S+G_{i}\overline{N_{1R}}\, L_{ia} \epsilon_{ab} D_{1b} +  P_{ik}\overline{N_{kR}}\, L_{ibL}\epsilon_{ab} D_{2b}
\nonumber\\
&+& M_1\overline{(N_{1 R})^c} N_{1 R} + M_{kl} \overline{(N_{Rk})^c} N_{Rl} +H.c.,
\end{eqnarray}

where $i,j=e,\mu,\tau$ and $k,l=2,3$, $a,b$ are $SU(2)$ indices, $\epsilon$ is the $SU(2)$ antisymmetric tensor.

The field transformations under the $B-L$ and discrete symmetries are given by the Table~\ref{table1} below.

\begin{table}[ht]
\centering
\begin{tabular}{|c|c|c|c|c|c|c|c|}\hline
& $L_i$ &  $l_{jR}$ & $N_{1 R}$ & $N_{2,3 R}$ & $S$ & $D_1$  & $D_2$ \\ \hline
$SU(2)_L\otimes U(1)_Y$ &  $(\textbf{2},-1)$ & $(\textbf{1},-2)$& $(\textbf{1},0)$ &$(\textbf{1},0)$ & $(\textbf{2},+1)$&$(\textbf{2},+1)$ & $(\textbf{2},+1)$\\ \hline
$B - L$  & - 1 & - 1  & $ - 5 $ &  + 4 & 0 & -4  & +5 \\ \hline
$Z_2$      & + & + & - & - & + & - & - \\ \hline
\end{tabular}
\caption{Transformation properties of the fermion and scalar fields under  $SU(2)_L\otimes U(1)_Y$, $  B - L $, and $\mathbb{Z}_3$. We do not include quarks because they are singlet under $Z_2$ and
$B-L=+1/3$ as usual.}
\label{table1}
\end{table}

The more general $SU(2)_L\otimes U(1)_{Y}$ and $\mathbb{Z}_2$ invariant
scalar potential for the three doublets, is given by:
\begin{eqnarray}
\label{potential}
V(S,D_1,D_2) & = & \mu^2_{SM} \vert S \vert^2 + \mu^2_{d1} \vert D_1 \vert^2  +\mu^2_{d2}  \vert D_2 \vert^2 +
\lambda_1 \vert S^\dag S \vert^2 + \lambda_2 \vert D_1^\dag
D_1\vert^2 \nonumber + \lambda_3 \vert D_2^\dag
D_2\vert^2 \nonumber \\ &+& \lambda_4 \vert S \vert^2\vert D_1 \vert^2 + \lambda_5 \vert S \vert^2\vert D_2\vert^2 + \lambda_6 \vert D_1 \vert^2\vert D_2\vert^2  + \lambda_7 \vert S^\dagger D_1 \vert^2  \nonumber \\ &+&
\lambda_8 \vert S^\dagger D_2 \vert^2 +[\mu^2_{12}D^\dagger_1 D_2+
\lambda_9(S^\dagger D_1)^2
+ \lambda_{10}(S^\dagger D_2)^2 + H.c.],
\end{eqnarray}


The quartic terms $\lambda_{9,10}$ break $B-L$ hardly. If the latter terms were zero the model has a global $B-L$ symmetry. For this reason they are naturally small \cite{tHooft:1979rat}. Hence, as noticed in Ref.~\cite{Merle:2015gea},
renormalisation group equations (RGEs) for those quantities will only allow for changes proportional to the couplings themselves, so that they remain small everywhere if they are small at any energy scale. Below, we will neglect these soft terms.

We will assume these parameters are real. Notice that if $\nu^2_{s(1,2)}S^\dagger D_{1,2}$ are allowed ($\mathbb{Z}_2$ is broken softly) and $\lambda_{9,10}=0$ the doublets are not inert anymore and we have a mechanism as in Ref.~\cite{Ma:2000cc} in which the smallness of the neutrino masses is due to the smallness of the VEV of the doublet(s) $D_1(D_2)$. On the other hand, if $\nu^2_{s1(2)}=0$ and $\lambda_{9,10}\not=0$ we have the scotogenic mechanism~\cite{Ma:2006km} with two inert doublet.  Here we will consider only the latter case.

Doing as usual the shifted as
$S^0~\!\!\!=\!\!\!~\frac{1}{\sqrt2}(v_{SM} + h + i G)$ with $v_{SM}=246$ GeV, and $D^0_{1,2}~=~
\frac{1}{\sqrt2}(R_{1,2}~+~iI_{1,2})$, the constraint equations are given by:
\begin{equation}
\label{vinculos}
v_{SM} (\mu^2_s  +   \lambda_1 v_{SM}^2)= 0.
\end{equation}

The masses matrices are all diagonal and the eigenvalues are:

For CP even scalars:
\begin{equation}
m^2_{h}  =  2 \lambda_1 v_{SM}^2, \;
m^2_{R1} =  \mu_{d1}^2 + \frac{v_{SM}^2}{2} (2 \lambda_{10} + \lambda_5 + \lambda_8),  \;
m^2_{R2} =  \mu_{d2}^2 + \frac{v_{SM}^2}{2} (2 \lambda_{9} + \lambda_5 + \lambda_8).
\label{cpp}
\end{equation}

For CP odd scalars:
\begin{equation}
m^2_{I1}  =  0, \;
m^2_{I2}  =  \mu_{d1}^2 + \frac{v_{SM}^2}{2} (- 2 \lambda_{10} + \lambda_5 + \lambda_8),  \;
m^2_{I3}  =  \mu_{d2}^2 + \frac{v_{SM}^2}{2} (- 2 \lambda_{9} + \lambda_5 + \lambda_8).
\label{cpi}
\end{equation}

For charged scalars:
\begin{equation}
m^2_{+1} = \frac{1}{4} ( 2 \mu_{d1}^2 +  \lambda_4 v_{SM}^2),\;
m^2_{+2}  =  \frac{1}{4} ( 2 \mu_{d2}^2 +  \lambda_5 v_{SM}^2).
\end{equation}
Nootice that, depending on the values of $\mu^2_d$ and the $\lambda$'s in Eqs.~(\ref{cpp}) and (\ref{cpi}) the neutral scalars may be lighter than the SM Higgs which mass is $m_h\approx125$ GeV.


The boundedness of the potential from below has to be a criteria defined allowing the greatest number of the parameter space. In the earlier work in Ref.~\cite{Chakrabortty:2013mha} was computed some set of vacuum stability conditions following 	the Copositive Criteria for the Boundedness of the Scalar Potential, in short the basic idea is construct the quartic couplings as a pure square of the combinations of bilinear scalar fields and set their coefficients could be non-negative, with this we can certainly makes the vacuum stable. However,  for scalar potential more complicated certain amount of ambiguities may arise, for more details see Refs.~\cite{Chakrabortty:2013mha, Kannike:2012pe}.

So in the base $\vert S\vert^2  , \vert D\vert^2, \vert D\vert^2 $, we have:
\begin{equation}
A=\left(\begin{array}{ccc}
\lambda_1 &  \lambda_4 & \lambda_5 + \lambda_9 r^2_{1}\\
\lambda_4  & \lambda_2 &  \lambda_6 + \lambda_{10} r^2_{2}\\
\lambda_5 + \lambda_9 r^2_{1} & \lambda_6 + \lambda_{10} r^2_{2} &  \lambda_3
\end{array}
\right).
\label{matrixa}
\end{equation}

The values for $r^2_i$ are those to minimize
the entries of the matrix. We have two relevant cases for the off-diagonal elements with sums: if both
coupling constants are positive/negative, the minimum
comes from choosing $r^2_i = 0$; if the constants have
opposite signs, the minimum comes from $r^2_i = 1$.

For a symmetric matrix A of order 3 the copositivity criteria are summarized as follows: $a_{ii} > 0$ and $v_{ij}= a_{ij} + \sqrt{a_{ii} a_{jj}} > 0$ and $\sqrt{a_{11}a_{22} a_{33}} + a_{12}\sqrt{a_{33}} +  a_{13}\sqrt{a_{22}} +  a_{23}\sqrt{a_{11}} + \sqrt{v_{12} v_{13} v_{23} } > 0$.
Explicitly we obtain:
\begin{eqnarray}
&&\lambda_1  >  0,
\quad
\lambda_2  >  0 ,
\quad
\lambda_3  >  0,
\quad
\lambda_4 + \sqrt{\lambda_1 \lambda_2}  >  0 ,
\nonumber  \\ &&
\lambda_5 - \lambda_9 + \sqrt{\lambda_1 \lambda_3}  >  0 ~~\text{or}, ~~  \lambda_5 + \sqrt{\lambda_1 \lambda_3}  >  0,
\nonumber  \\ &&
\lambda_6 - \lambda_{10} + \sqrt{\lambda_2 \lambda_3}  >  0 ~~\text{or}, ~~
\lambda_6 + \sqrt{\lambda_2 \lambda_3}  >  0
\label{nossa1}
\end{eqnarray}
and
\begin{eqnarray}
\sqrt{(\lambda_1 \lambda_2 \lambda_3)} + \lambda_{4} \sqrt{\lambda_3}
+
(\lambda_5 + \lambda_9) \sqrt{\lambda_2} + (\lambda_6 + \lambda_{10}) \sqrt{\lambda_3}  > 0,
\label{nossa2}
\end{eqnarray}

It is easy to verify that if the constraints in Eqs.~(\ref{nossa1}) are satisfied the conditions in Eq.~(\ref{nossa2}) are automatically satisfied. Hence, the positivity of the scalar potential is guarantee just by the conditions in Eq.~(\ref{nossa1}).
Moreover, the scalar has a global minimum since
\begin{equation}
V\left(\frac{v_{SM}}{\sqrt2},0,0\right)=-\lambda_1v^4_{SM}.
\label{msp}
\end{equation}

\section{Lepton masses and mixing}
\label{sec:leptons}

\begin{center}
	\begin{figure}[!ht]
		\subfloat[]{\includegraphics[width=6.5cm]{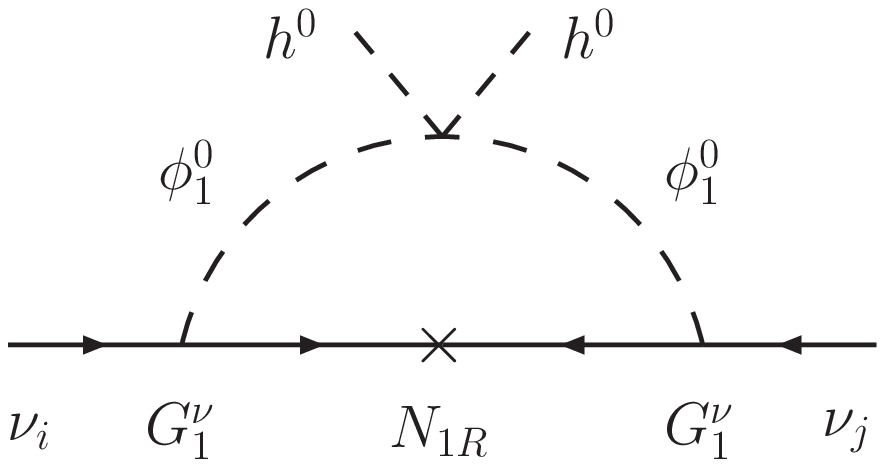}} \qquad
		\subfloat[]{\includegraphics[width=6.5cm]{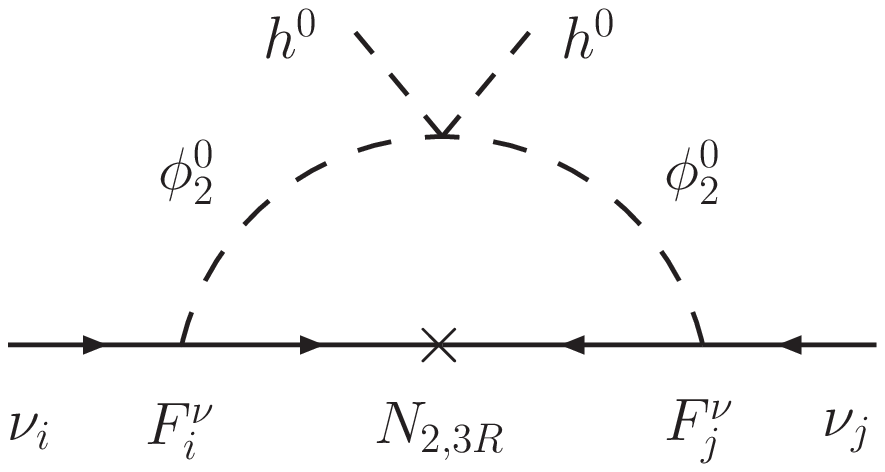}}
		\caption{ Loops diagrams for neutrinos masses originated from Eqs.~(\ref{yukawa}) and (\ref{potential}).}\label{loops}	
	\end{figure}
\end{center}

Notice that neutrinos are still massless at tree level as in SM. However, in the scalar potential there are the interactions like $\lambda_4$ and $\lambda_5$ in the scalar potential in (\ref{potential}),
that induce the sort of diagrams as those in Fig.~\ref{loops}. With these interactions, it is possible to implement the mechanism of Ref.~\cite{Ma:2006km} for the radiative generation of neutrinos masses.

In fact, the diagram in Fig.~\ref{loops} are exactly calculable from the exchange of $\textrm{Re} D^0_{1,2}$ and $\textrm{Im} \phi^0_{1,2}$~\cite{Ma:2006km}
\begin{eqnarray}
(M^\nu)_{ij} &=&  \frac{G_{i}G_{j} M_{1}}{32\pi^2} \left[ \frac{m^2_{R1}}{m^2_{R1}-M^2_1} \ln \frac{m^2_{R1}}{M_{1}^2}- \frac{m^2_{I1}}{m^2_{I1}-M^2_1} \ln \frac{m_{I1}^2}{M_{1}^2}\right]
\nonumber \\ &+&\frac{P_{ik}P_{jk}  M_{k}}{32\pi^2} \left[ \frac{m^2_{R2}}{m^2_{R2}-M^2_k} \ln \frac{m^2_{R2}}{M_{k}^2}- \frac{m^2_{I2}}{m^2_{I2}-M^2_k} \ln \frac{m_{I2}^2}{M_{k}^2}\right],
\label{numass12}
\end{eqnarray}
where $m_{Ra}$ and $m_{Ia}$ with $a=1,2$ are the masses of $\textrm{Re}D^0_{1,2}$ and $\textrm{Im}D^0_{1,2}$, respectively.
We can define  $\Delta^2_1= m^2_{R1}-~m^2_{I1}=~2\lambda_{10} v_{SM}^2 $,  $\Delta^2_2= m^2_{R2}-~m^2_{I2}=~2\lambda_{9} v_{SM}^2 $, and $m^2_{0a}=(m^2_{Ra}+m^2_{Ia})/2$, $a=1,2$. If $\lambda_{9,10}\ll1$, we can write
\begin{eqnarray}
(M^\nu)_{ij} \!\!&=&\!\! \frac{v_{SM}^2}{8 \pi^2} \left[\lambda_{10} \frac{G_{i} G_{j} M_1}{m^2_{01} - M_1^2} \left(1 - \frac{M_1^2}{m^2_{01} - M_1^2} \ln\frac{m^2_{01}}{M^2_1}\right)\right.\nonumber \\& +& \left.\lambda_{9} \frac{P_{ik} P_{jk} M_k}{m^2_{02} - M_k^2} \left(1 - \frac{M_k^2}{m^2_{02} - M_k^2} \ln\frac{m^2_{02}}{M^2_k} \right) \right],
\label{numass2}
\end{eqnarray}
where we have omitted a sum in $k=2,3$.

In order to obtain the active neutrinos masses we assume a normal hierarchy and, without loss of generality, that $M_1 \sim M_2\sim M_3$ and will be represented from now on by $M_R$. $M^\nu$ is diagonalized with a unitary matrix $V_L^\nu$ i.e.,  $\hat{M}^\nu~=~V^{\nu T}_LM^\nu V^\nu_L$,
where $\hat{M}^\nu = \textrm{diag} (m_1, m_2, m_3)$, just for simplicity we assume here $\hat{M}^\nu = \textrm{diag} (0 , 0.0086, 0.05)$ eV.

It is important to note from these considerations, that there exist a multitude of other possibilities which satisfy also the masses squared differences and the astrophysical limits in the active neutrino sector. Each one corresponds to different parametrization of the unitary matrices $V^l_{L,R},V^\nu_L$.

To obtain the neutrinos masses from Eq.~(\ref{numass2}), we have as free parameters $\lambda_9,\lambda_{10}$, $M_R$, $m^2_{01}$, $m^2_{02}$ and the Yukawa couplings. For the sake of simplicity we sue $\lambda_9=\lambda_{10}\equiv \lambda$ and  $m^2_{01}=m^2_{02}\equiv m_0$.

\begin{table}[!h]
\begin{tabular}{|c|c|c|c|c|c|c|c|}\hline
Sol. &
Masses in TeV & $G^l_{11} $  & $G^l_{12}$ & $G^l_{13}$  & $G^l_{22}$ & $G^l_{23} $ &  $G^l_{33}$  \\ \hline
P1 & $M_R=0.5 $, $m_0 = 2.2 $
& 0.000421836  &  0.000514741 & -0.000800772 &   0.00374767 & -0.00356758 & 0.00367645   \\ \hline
P2 & $M_R=1.5\ \ $, $m_0 = 2.2 $
& 0.000421836  &  0.000514741 & -0.000800772 &   0.00374767 & -0.00356758 &  0.00367641  \\ \hline
P3 & $M_R=2.5 $, $m_0 = 2.2 $
& 0.000421836  &  0.000514741& -0.000800772 &   0.00374767 & -0.00356758&  0.00367645  \\ \hline
P4 & $M_R=3\ \ $, $m_0 = 2.2 $
& 0.00042195  &  0.000515086 & -0.000801103 &   0.00374772 & -0.00356751 & 0.00367645  \\ \hline
\end{tabular}
\caption{Masses of the scalars in Scotogenic model (in GeV).}
\label{table4}
\end{table}

The mass matrices in the charged sector $M^l$ are diagonalized by a bi-unitary transformation $\hat{M}^l=V^{l\dagger}_L M^l V^l_R$ and $\hat{M}^l = \textrm{diag} (m_e, m_\mu, m_\tau)$. The relation between symmetry eigenstates (primed) and mass (unprimed) fields are $l^\prime_{L,R}=V^l_{L,R}l_{L,R}$
and $\nu^\prime_L=V^\nu_L \nu_L$, where
$l^\prime_{L,R}=(e^\prime,\mu^\prime,\tau^\prime)^T_{L,R}$, $l_{L,R}=(e,\mu,\tau)^T_{L,R}$  and
$\nu ^\prime_L=(\nu_e,\nu_\mu,\nu_\tau)^T_L$
and $\nu_L=(\nu_1,\nu_2,\nu_3)_L$.
Defining the lepton mixing matrix as $V_{PMNS}=V^{l\dagger}_LV^\nu_L$, it means that this matrix appears in the charged currents coupled to $W_\mu^+$. We obtain:

\begin{equation}
\vert V_{PMNS}\vert \approx\left(\begin{array}{ccc}
0.815 & 0.565 & 0.132\\
0.479 & 0.527 & 0.702\\
0.327 & 0.635 & 0.700\\
\end{array}\right),
\label{pmns}
\end{equation}
which is in agreement within the experimental error data at 3$\sigma$ given by~\cite{GonzalezGarcia:2012sz}
\begin{equation}
\vert V_{PMNS}\vert \approx\left(\begin{array}{ccc}
0.795-0.846& 0.513-0.585 & 0.126-0.178\\
0.4205-0.543 & 0.416-0.730  & 0.579 - 0.808 \\
0.215 - 0.548 & 0.409 - 0.725 & 0.567 -0.800 \\
\end{array}\right),
\label{pmnsexp}
\end{equation}
and we see that it is possible to accommodate all lepton masses and the PMNS matrix. Here we do not
consider $CP$ violation.

\section{phenomenological results}
\label{sec:pheno}

\begin{figure}[!h]
\subfloat[]{\includegraphics[width=5.5cm]{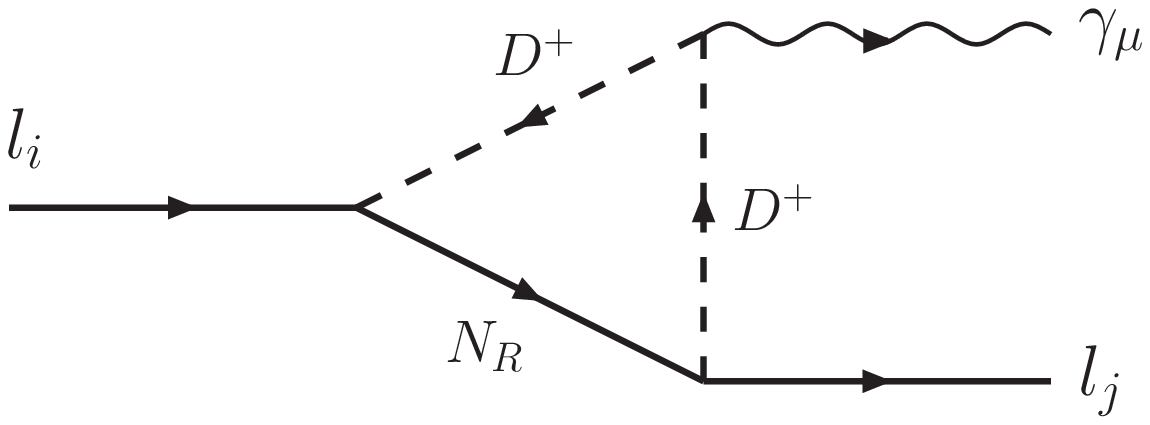}} \qquad
\subfloat[]{\includegraphics[width=5.5cm]{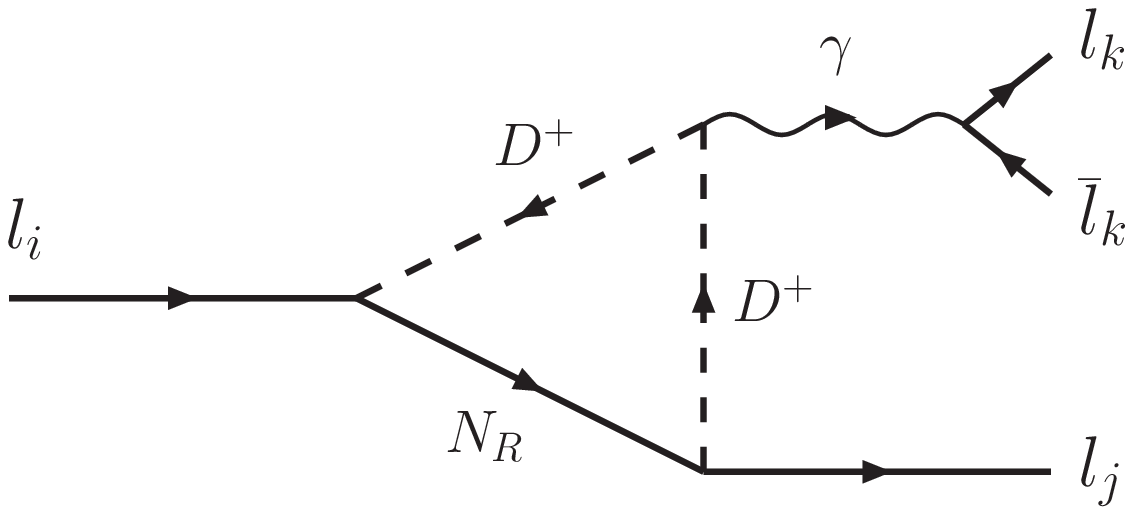}}
\caption{Decays (a) $l_i\to l_j\gamma$ and (b) $l_i\to l_jl_k\bar{l}_k$ in the B-L model. Generic sample contribution from a sterile neutrino $N_R$ and a charged scalar $D^+$, also bubble diagrams contribute.}
\label{FIGURE-loops}
\end{figure}

Considering the Yukawa values derived in the Sec. \ref{sec:leptons} we can analyze the decays $\mu \to e \gamma$ in Fig~\ref{FIGURE-loops}~(a) and $\mu \to eee$ in \ref{FIGURE-loops}~(b), as function of the sterile neutrino mass $m_N  [100, 4000]$ GeV with given values for the charged scalar $m_{D^+}$ = 80, 250, 500, 750 GeV. We have tested the five different parameterizations of the $V^l_L$ matrices derived in the Sec. \ref{sec:leptons} (see Table \ref{table4}).
Lepton flavor violation (LFV) processes will be induced by the existence of Yukawa interaction $G_{i} L_i \epsilon N D_1$ and $P_{ik} L_i \epsilon N_k D_2;\;i=e,\mu,\tau,\;k=2,3$. As a check for this model we will first take in account the current
most stringent one, i.e., the MEG experiment on the radiative decay
$\mu \to e\gamma$ with BR$(\mu\to e \gamma)<4.2\times10^{-13}$~\cite{TheMEG:2016wtm}.

Our obtained space of parameters of the model allow us to identify that from the different processes $l_i\to l_j\gamma$, $l_i\to l_jl_k\bar{l}_k$ and $g-2$, it is the $\mu\to e\gamma$ decay that imposes the most stringent bounds on the mass of the $N_R$ right-handed neutrino. For the $\mu\to e\gamma$ channel our obtained space of parameters has room for predicting signals between the current experimental upper limit and the expected upcoming one, see Table~\ref{Table-loop-experiments}, therefore we are going to focus in this interval.

\begin{figure}[!h]
\includegraphics[width=18cm]{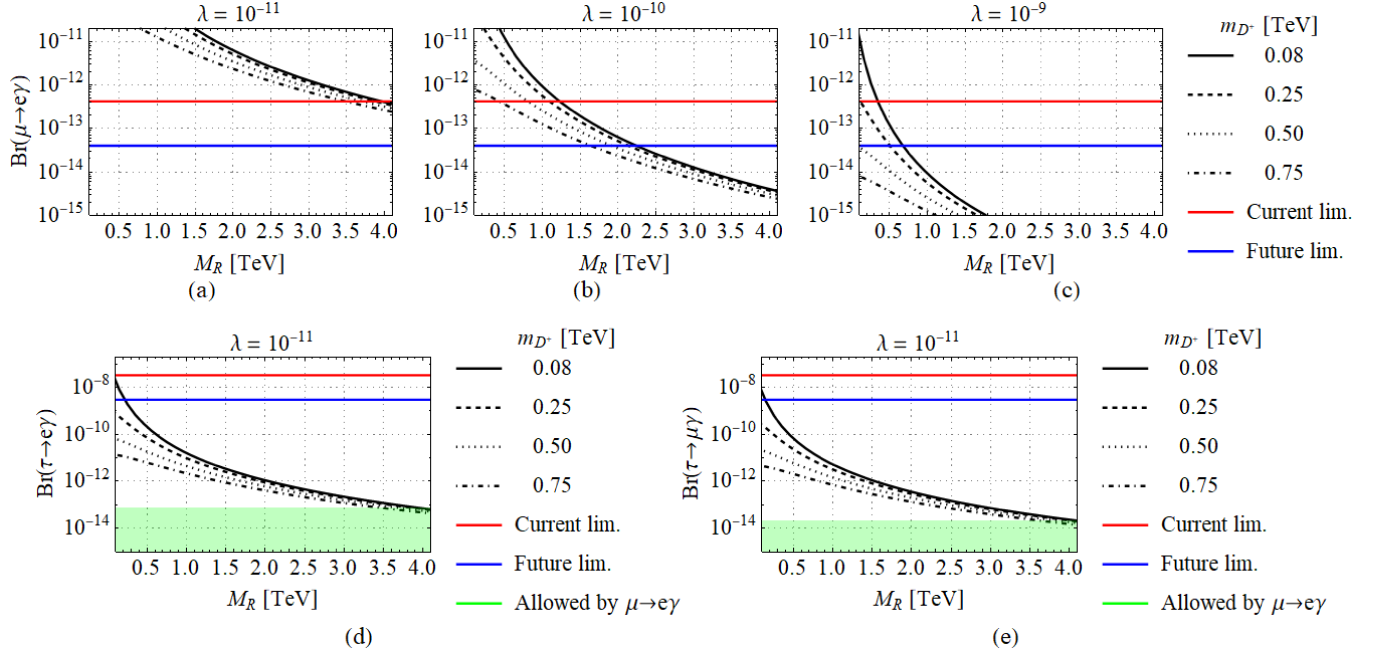}
\caption{Br$(l_i\to l_j\gamma)$ as function of $M_R=[0.1,4]$ TeV with fixed $\lambda$ and $m_{D^+}$.
(a)-(c) Br($\mu\to e\gamma$) demands the mass constraints listed in the Table~\ref{TABLE-mass-bounds},
hence $\mu\to e\gamma$ allows (d) $\text{Br}(\tau\to e\gamma)$ $<$ $7.39\times10^{-14}$ and
(e) $\text{Br}(\tau\to \mu\gamma)$ $<$ $2.21\times10^{-14}$.}\label{FIGURE-li-Alj-B-L}
\end{figure}

The new particle content of the model that induces $\mu\to e\gamma$ is $D_1^+ \& N_2$, $D_2^+ \& N_3$, $D_1^+\& N_s$ and $D_2^+\& N_s$, see the Fig.~\ref{FIGURE-loops}(a).
The $\lambda$ parameter dependent of the Yukawas $G_{il}$
has the wide interval $\lambda=[10^{-11},10^{-1}]$, from which the gap $\lambda=[10^{-11},10^{-9}]$ allows the
$\mu\to e\gamma$ signals showed in the Figs.~\ref{FIGURE-li-Alj-B-L}(a)-(c) for the given scenarios of $m_{D^+}$. The current experimental upper limit, depicted with the red line, demands to set bounds to the sterile neutrino mass in order to respect it, such bounds are listed in the Table~\ref{TABLE-mass-bounds}. From $\lambda>10^{-9}$  the $\mu\to e\gamma$ signal is suppressed away from the current experimental upper limit.
For the $\tau\to l_j\gamma$ channels, Figs.~\ref{FIGURE-li-Alj-B-L}(d)-(e), we predict (d) $\text{Br}(\tau\to e\gamma)$ $<$ $7.39\times10^{-14}$ and
(e) $\text{Br}(\tau\to \mu\gamma)$ $<$ $2.21\times10^{-14}$ allowed by $\mu\to e\gamma$.

\begin{figure}[!h]
\includegraphics[width=18cm]{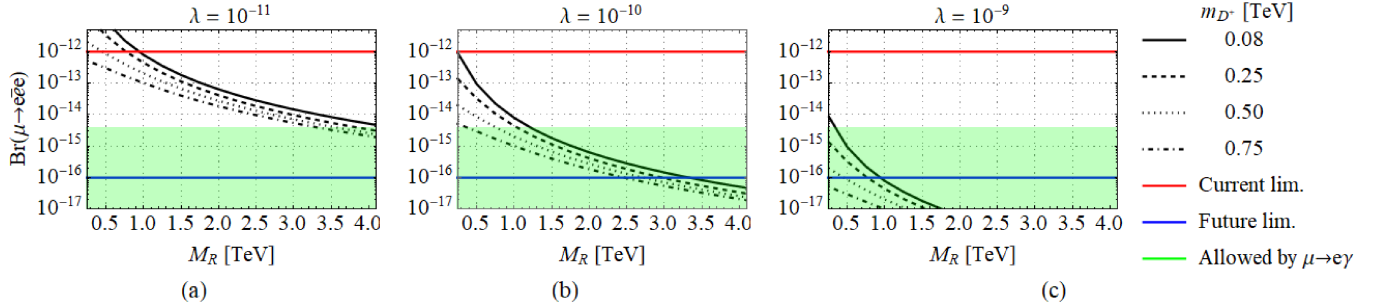}
\caption{$\text{Br}(\mu\to ee\bar{e})$ as function of $M_R=[0.1,4]$ TeV, with fixed $\lambda$ and $m_{D^+}$.
$\text{Br}(\mu\to ee\bar{e})$ $\leq$ $3.7\times10^{-15}$ allowed by $\mu\to e\gamma$.}
\label{FIGURE-li-ljlklk-B-L}
\end{figure}

The $l_i\to l_jl_k\bar{l}_k$ decay arises when the $\gamma l_k\bar{l}_k$ vertex is attached to the photon in $l_i\to l_j\gamma$, see Fig.~\ref{FIGURE-loops}(b). We have found that $\text{Br}(\mu\to ee\bar{e})$ $<$ $3.7\times10^{-15}$ allowed by $\mu\to e\gamma$, see Fig.~\ref{FIGURE-li-ljlklk-B-L}, this prediction is quite interesting because it starts one order of magnitude above the expected upcoming experimental upper limit. Regarding to the tau decays into three bodies, they are far from its corresponding experimental upper limit, therefore we do not present them.

\begin{figure}[!h]
\includegraphics[width=4cm]{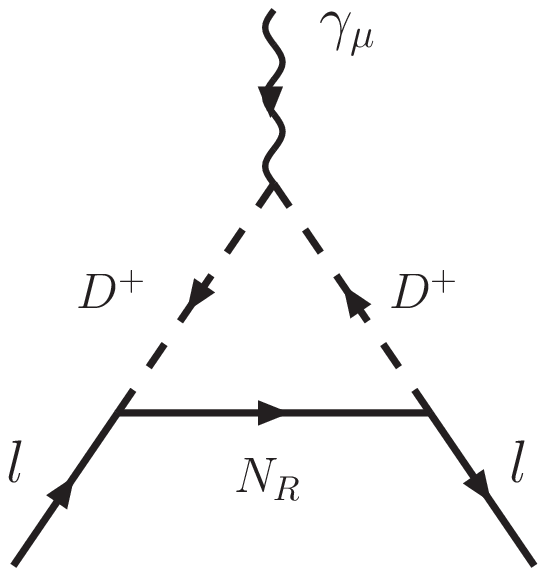}
\caption{$g-2$ in the B-L model.}
\label{FIGURE-g-2-B-L-Diagram}
\end{figure}

\begin{figure}[!h]
\includegraphics[width=18cm]{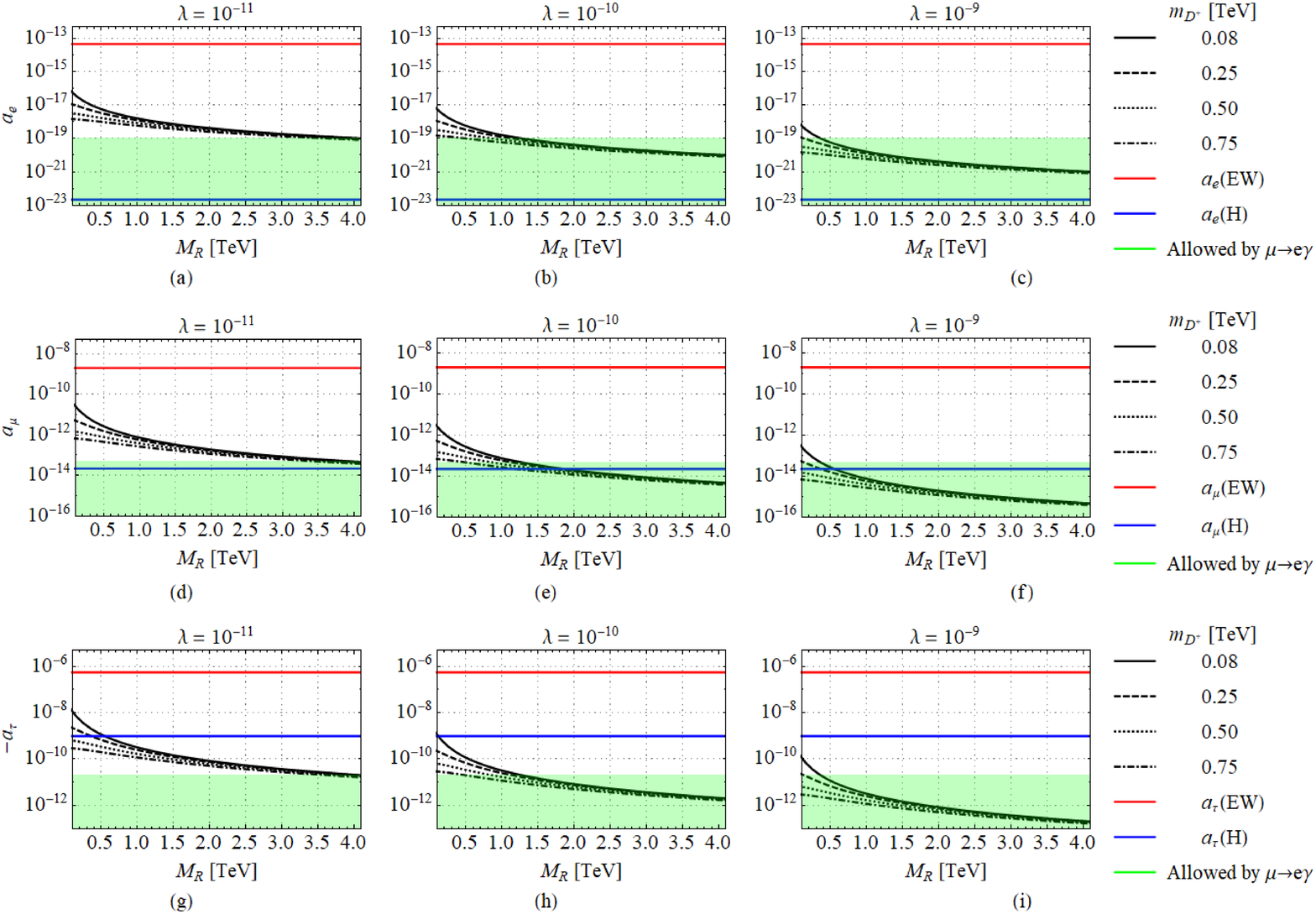}
\caption{Anomalous magnetic dipole moments as function of $M_R=[0.1,4]$ TeV with fixed $m_{D^+}$ and $\lambda$.
The $\mu\to e\gamma$ decay allows: (a)-(c) $a_e<1.08\times10^{-19}$,
(d)-(f) $a_\mu<4.88\times10^{-14}$,
(g)-(i)~ $a_\tau<-2.16\times10^{-11}$.
As references, the horizontal red line corresponds to the $a_l(\text{EW})=a_l(W)+a_l(Z)$ the SM electroweak contribution and the blue one to the $a_l(H)$ the Higgs boson contribution.}
\label{FIGURE-magnetic-dipoles}
\end{figure}

For completeness we also present the effect of the B-L model on the anomalous magnetic dipole moments of charged leptons, see Fig.~\ref{FIGURE-g-2-B-L-Diagram}. The $\mu\to e\gamma$ allows to predict the signals given in Fig.~\ref{FIGURE-magnetic-dipoles}, where as reference values we use $a_{l}(\text{EW})=a_{l}(W)+a_{l}(Z)$ the SM electroweak signal depicted with the red line and also $a_l(H)$ the Higgs boson contribution with a blue line. Specifically,
in the Figs.~\ref{FIGURE-magnetic-dipoles} our prediction are: (a)-(c) $a_e<1.08\times10^{-19}$,
(d)-(f) $a_\mu<4.88\times10^{-14}$,
(g)-(i)~ $a_\tau<-2.16\times10^{-11}$.

\begin{table}[ht]
  \centering
\begin{tabular}{|c|c|c|c|c|}\hline
Decay                  & Current limit         & Future limit         & SM \\
\hline
Br($\mu\to e\gamma$)   & $< 4.2\times10^{-13}$ \cite{TheMEG:2016wtm} & $<6.0\times10^{-14}$ \cite{Baldini:2013ke}
 & $10^{-48}$ \\
Br($\tau\to e\gamma$)  & $< 3.3\times10^{-8\ }$ \cite{Olive:2016xmw} & $<3.3\times10^{-9\ }$ \cite{Aushev:2010bq} & $10^{-49}$ \\
Br($\tau\to\mu\gamma$) & $< 4.4\times10^{-8\ }$ \cite{Olive:2016xmw} & $<3.3\times10^{-9\ }$ \cite{Aushev:2010bq} & $10^{-49}$ \\
\hline
\end{tabular}
\caption{$l_i\to l_j\gamma$, experimental upper limit and the SM predictions.}\label{Table-loop-experiments}
\end{table}

\begin{table}[ht]
\begin{tabular}{|c|c|c|c|}
\hline
\multicolumn{4}{|c|}{$\mu\to e\gamma$} \\
\cline{1-4}
\multirow{2}{*}{$m_{D^+}$ [GeV]} & \multicolumn{3}{c|}{$M_R$ [GeV]} \\
  \cline{2-4}
& $\lambda=10^{-11}$   & $\lambda=10^{-10}$ & $\lambda=10^{-9}$  \\
\hline
~80  & $>$ 3961 & $>$ 1227 & $>$ 343 \\
250  & $>$ 3882 & $>$ 1090 & $>$ 109 \\
500  & $>$ 3706 & $>$ ~807 & $-$    \\
750  & $>$ 3486 & $>$ ~427 & $-$    \\
\hline
\end{tabular}
\caption{\ref{FIGURE-li-Alj-B-L}
Mass constraints for the right-handed neutrino, with fixed $m_{D^+}$ and $\lambda$, in order to respect
Br$(\mu\to e\gamma)^\text{Exp}<4.2\times10^{-13}$.}\label{TABLE-mass-bounds}
\end{table}

\section{The scotogenic mechanism in a susy extension of the model}
\label{sec:susymodel}

Let us consider the supersymmetric version of this model. We show the details only of the lepton sector. More details of this model will be presented elsewhere. For the moment let us consider just the lepton sector
and their chiral superfields with the transformation properties shown in Table~\ref{leptonssusy}.

\begin{table}[h]
	\begin{center}
		\begin{tabular}{|c|c|c|c|c|}
			\hline
			${\rm{Superfield}} $ & $\hat{L}_{iL}$ & $\hat{e}_{iR}$ & $\hat{N}_{1R}$ & $\hat{N}_{\beta R}$ \\
			\hline
			$(SU(2)_{L},U(1)_{Y})$ & $( {\bf 2},-1)$ & $( {\bf 1},2)$ & $( {\bf 1},0)$ & $( {\bf 1},0)$ \\
			\hline
			$B-L$         & $-1$ & $+1$ & $5$ & $-4$ \\ \hline
			$\mathbb{Z}_2$ & +    & +    & -   & -\\
			\hline
		\end{tabular}
	\end{center}
\caption{Transformation properties of the fermions under $SU(2)_{L}\otimes U(1)_{Y}$ and $B-L$. The generation indices are as follows: $i=1,2,3$ and
$\beta =2,3$.}
\label{leptonssusy}
\end{table}

The new scalars, the sneutrinos, are innert due $B-L$ symmetry,
therefore, we can write
\begin{equation}
\langle \tilde{\nu}_{iL} \rangle =
\langle \tilde{N}_{iR} \rangle =  0.
\label{vevextradoublets}
\end{equation}

In the scalar sector, as usual the supersymmetric partner of the SM scalar doublet (here we change the notation $S\to H_1$) is denoted by $H_2$, moreover in this case we have to introduce more scalar fields, the doublets $D^\prime_{1,2}$, and the singlets $\varphi$ and $\phi$ with the quantum numbers shown in Table~\ref{scalarssusy}.
\begin{table}[h]
	\begin{center}
		\begin{tabular}{|c|c|c|c|c|c|c|c|c|}
			\hline
			 & $H_{1}$ & $H_{2}$ & $D_{1}$ & $D^{\prime}_{1}$ & $D_{2}$ & $D^{\prime}_{2}$ &
			$\varphi$ & $\phi$ \\
			\hline
			$(SU(2)_{L},U(1)_{Y})$ & $( {\bf 2},1)$ & $( {\bf 2},-1)$ &
			$( {\bf 2},1)$ & $( {\bf 2},-1)$ &
			$( {\bf 2},1)$ & $( {\bf 2},-1)$ & $( {\bf 1},0)$ & $( {\bf 1},0)$ \\
			\hline
			$B-L$          & $0$ & $0$ & $-4$ & $4$ & $5$ & $-5$ & $-10$ & $8$ \\ \hline
			$\mathbb{Z}_2$ & + & +  & - & - & - & - & + & +\\
 			\hline
\end{tabular}
\end{center}
\caption{Transformation properties of the scalar fields under the $SU(2)_{L}\otimes U(1)_{Y}$ and $B-L$ symmetries.}
\label{scalarssusy}
\end{table}

The usual scalars $H_{1,2}$ their VEVs as usual are given as usual $ \langle H_{1} \rangle = v_1/\sqrt{2}$, $
\langle H_{2} \rangle = v_2/\sqrt{2}$, while the new scalars $D_{1},D^{\prime}_{1},D_{2},D^{\prime}_{2}$ are innert due
$B-L$ symmetry, $\langle D_{1} \rangle = \langle D_{2} \rangle =
\langle D^{\prime}_{1} \rangle =
\langle D^{\prime}_{2} \rangle = 0. $
The VEV of the new scalars in singlets $\langle \varphi \rangle = u_1/\sqrt{2}$, and $
\langle \phi \rangle = u_2/\sqrt{2}$. Therefore, as in
the Minimal Supersymmetric Standard Model (MSSM) the charged
$W$-boson has mass given by
$M_{W}=(gv_2/2) \sqrt{1+ \tan^{2}\beta}$ where
$\tan \beta =(v_1/v_2)$. Terms like $\lambda_{9,10}$ which appear in Eqs.~(\ref{potential}), do not appear in the susy version.

The interactions in the superpotential that are important for the generation of the neutrino masses are
\begin{eqnarray}
{\cal L}^{leptons}_{Yukawa}&=&
G^{l}_{ij}\bar{L}_{iL} l_{jR}
H_2+G_{i}L^T_{iaL}\epsilon_{ab} C^{-1}N_{1R}D_{1b}+
P^{\nu}_{ik}L^T_{iaL} C^{-1}\epsilon_{ab}N_{k R}D_{2b}\nonumber \\ &+&
H_{11} \overline{(N_{1R})^c} N_{1R}\,\varphi+
H_{kl }\overline{(N_{k R})^c}N_{l R}\,\phi,
\nonumber \\
\label{yukawasusy1}
\end{eqnarray}
and the last two terms above will
generate Majorana mass terms to the right-handed
neutrinos when the scalars $\varphi$ and $\phi$ get an non-zero VEV.

Therefore three neutrinos get mass at tree level and
we have three massless neutrinos, in the same way as we
presented above.

In our superpotential we need to break the $B-L$ symmetry and it is generate the following interactions
\begin{eqnarray}
{\cal L}^{break}_{B-L}&=&\lambda_{ijk}\left[
\tilde{\nu}^{i}_{L}\bar{l}^{k}_{R}l^{j}_{L}+
\tilde{l}^{j}_{L}\bar{l}^{k}_{R}\nu^{i}_{L}+
(\tilde{l}^{k}_{R})^{*}(\bar{\nu}^{i}_{L})^{c}l^{j}_{L}-
\left( i \longleftrightarrow j \right) + H.c.
\right].
\label{yukawasusy2}
\end{eqnarray}
The couplings $\lambda_{ijk}$ can contribute to various (low-energy)
process: charged current universality, bound on masses of
$\nu_{e, \mu , \tau}$ and etc, for more details about this
subject see Ref.~\cite{Barbier:2004ez,dress,Baer:2006rs}.

\begin{center}
\begin{figure}[ht]
\subfloat[]{\includegraphics[width=6.5cm]{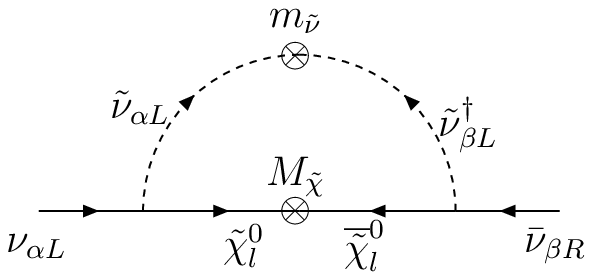}} \qquad
\subfloat[]{\includegraphics[width=6.5cm]{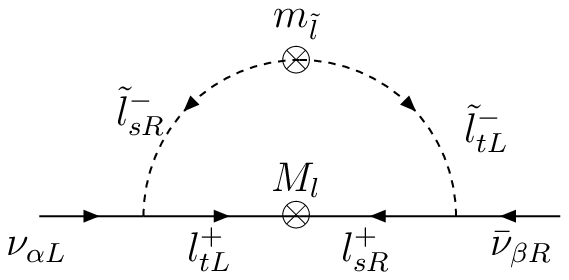}}
\caption{The one loop correction to the masses of $m_{\nu_{\alpha}}$ in the susy version of the model. The neutralino-neutrino-sneutrino contributions are in (a), while the sleptons contributions in (b). The respective vertices are given in the text.}
\label{figsusy}
\end{figure}
\end{center}

The quartic interactions which appear in the diagram Fig.~\ref{figsusy} is
the coefficients of various quartic interaction are
\begin{eqnarray}
&& D[v_2,v_2, \tilde{\nu}_{i}, \tilde{\nu_{j}}]=-d_{g}[ \tilde{\nu}]
c_{2 \alpha}\delta_{ij}, \,\
D[v_1,v_2, \tilde{\nu}_{i}, \tilde{\nu_{j}}]=2d_{g}[ \tilde{\nu}]
s_{2 \alpha}\delta_{ij} , \nonumber \\&&
D[v_1,v_1, \tilde{\nu}_{i}, \tilde{\nu_{j}}]=d_{g}[ \tilde{\nu}]
c_{2 \alpha}\delta_{ij},
\end{eqnarray}
and
\begin{eqnarray}
&& D[v_2,v_2, \tilde{l}_{s}, \tilde{l_{t}}]=-
d_{Y}[ \tilde{l}_{s}, \tilde{l_{t}}]c^{2}_{ \alpha}-d_g[\tilde{l}_s,\tilde{l}_t]c_{2\alpha},\nonumber \\&&
D[v_1,v_2, \tilde{l}_{s}, \tilde{l_{t}}]=
d_{Y}[ \tilde{l}_{s}, \tilde{l_{t}}]s_{2 \alpha}+2
d_{g}[ \tilde{l}_{s}, \tilde{l_{t}}]s_{2 \alpha}, \nonumber \\&&
D[v_1,v_1, \tilde{l}_{s}, \tilde{l_{t}}]=-
d_{Y}[ \tilde{l}_{s}, \tilde{l_{t}}]s^{2}_{ \alpha}+
d_{g}[ \tilde{l}_{s}, \tilde{l_{t}}]c_{2 \alpha}, \,\
\end{eqnarray}
we omitted the $v_iv_j$ factors, and we have used the notation~\cite{dress}:
\begin{eqnarray}
d_{g}[ \tilde{\nu}]&=& \frac{g^{2}}{8} \left( 1+t^{2}_{W} \right), \,\
d_{g}[ \tilde{l}_{s}, \tilde{l_{t}}]=
\frac{g^{2}}{4M^{2}_{W}c^{2}_{\beta}}m^{2}_{\tilde{f}}\,
c_{( \theta_{\tilde{s}}- \theta_{\tilde{t}})}, \nonumber \\
d_{Y}[ \tilde{l}_{s}, \tilde{l_{t}}]&=&- \frac{g^{2}}{8} \left[
2t^{2}_W  s_{\theta_{\tilde s}} s_{\theta_{\tilde t}}
+
c_{\theta_{\tilde s}} c_{ \theta_{\tilde t}}
\left( 1-t^{2}_{W} \right)
\right], \nonumber \\
\end{eqnarray}
where $\theta_{\tilde{f}}$ is the mixing angle in the charged
slepton sector, $m^{2}_{\tilde{f}}$ their mass, while the symbols
to Weinberg angle $\theta_{W}$ are $s_{W},c_{W},t_{W}$.
where the rotations angle $\alpha$ and $\beta$ seen to obey
the relations \cite{dress,Baer:2006rs}
\begin{eqnarray}
s_{2 \alpha}&=&-
\frac{M^{2}_{H^{0}}+M^{2}_{h^{0}}}{M^{2}_{H^{0}}-M^{2}_{h^{0}}}s_{2 \beta},
\,\
c_{2 \alpha}=-
\frac{M^{2}_{A^{0}}-M^{2}_{Z}}{M^{2}_{H^{0}}-M^{2}_{h^{0}}}c_{2 \beta},
\nonumber \\
t_{2 \alpha}&=&
\frac{M^{2}_{H^{0}}+M^{2}_{h^{0}}}{M^{2}_{A^{0}}-M^{2}_{Z}}
t_{2 \beta},
\end{eqnarray}
where $h$ is the lighest CP-even Higgs, $H$ the heavy CP-even
Higgs and $A$ is the pseudo-scalar of MSSM. For any angle $\zeta$, we use $s_{\zeta},c_{\zeta},t_{\zeta}$
to mean $\sin \zeta$, $\cos \zeta$ and $\tan \zeta$ respectively. Finally, the neutrino-neutralino-sneutriono interaction $(\bar{\nu}_{iL}\tilde{\chi}^0_l\tilde{\nu}^*)$, (up to a factor $-i/\sqrt{2}$) is given by  $gZ^*_{l1}s_W+g^\prime Z^*_{l2}c_W$.

Take into account this fact together with Eq.~(\ref{yukawasusy1}) and the quartic interactions which are proportional to $v^2_1,v^2_2,v_1v_2$, we get the following one loop correction, see Fig.~\ref{fig1}, to the neutrinos masses (assuming that $v_1\gg v_2$)
\begin{eqnarray}
(M^\nu)_{ij}&=& (M^\nu)^{NS}_{ij}+\frac{
	\left( gZ^{\ast}_{l1}s_W+g^{\prime}Z^{\ast}_{l2}
	c_W\right)^2 }{64 \pi^{2}}\nonumber \\ &\cdot&M_{\tilde{\chi}^{0}_{l}}
\left[
\frac{m^{2}_{\tilde{\nu_{l_{s}}}}}{m^{2}_{\tilde{\nu_{l_{s}}}}-M^{2}_{\tilde{\chi}^{0}_{l}}}
\ln\left( \frac{m^{2}_{\tilde{\nu_{l_{s}}}}}{M^{2}_{\tilde{\chi}^{0}_{l}}}  \right)-
\frac{M^{2}_{\tilde{\nu_{l_{t}}}}}{M^{2}_{\tilde{\nu_{l_{t}}}}-m^{2}_{\tilde{\chi}}}
\ln\left( \frac{m^{2}_{\tilde{\nu_{l_{t}}}}}{M^{2}_{\tilde{\chi}^{0}_{l}}}  \right)  \right]
\nonumber \\ &+&
M_l\,\frac{\lambda_{ist}\lambda_{jst}}{16 \pi^{2}}
s^2_{(\theta_{\tilde{l}_{s}}+ \theta_{\tilde{l}_{t}})}
\left[
\frac{m^{2}_{\tilde{l}_{s}}}{m^{2}_{\tilde{l}_{s}}-M^{2}_l}
\ln\left( \frac{m^{2}_{\tilde{l}_{s}}}{M^{2}_l }  \right)-
\frac{m^{2}_{\tilde{l}_{t}}}{m^{2}_{\tilde{l}_{t}}-M^{2}_l}
\ln\left( \frac{m^{2}_{\tilde{l}_{t}}}{M^{2}_l}  \right)  \right],
\end{eqnarray}
where $(M^\nu)^{NS}$ denote the non-susy contributions given in Eq.~(11), $Z_{lk}$ denotes the mixing matrix in the neutralino sector, $m_{\tilde{\nu_l}} $ denotes the mass of the sneutrinos,
$M_{\tilde{\chi}^{0}_{l}}$ the mass of the
neutralinos, while
$M_l$ is the mass of the exchanged lepton and $m_{\tilde{l}_{s}}$
is the mass of their respective slepton. Notice that the contibutions of  neutralinos are larger that those of the charged leptons since $M_{\tilde{\chi}^{0}_{l}}\gg M_l$. If we out $(M^\nu)^{NS}=0$, chosing $\lambda_9=\lambda_{10}=0$ the numerical results obtained in Sec.~\ref{sec:leptons} can be attributed to the neutralino contributions.

There is also contributions for the neutrino masses at 1-loop coming from the interactions $\hat{L}\hat{Q}\hat{d}$, however we can forbid them assuming a discrete symmetry $\mathbb{Z}_3$ under which $\hat{Q}\to \omega^\prime \hat{Q}$
$\hat{u}\to \omega^{\prime -1}\hat{u}, \hat{d}\to \omega^{\prime -1}\hat{d}$, with all other fields being even under
this transformation. However, these extra contributions to the neutrino mass could be taken into account

The interactions in Eq.~(\ref{yukawasusy2}) also induce flavour
changing neutral currents, for instance, will generater the
following decay:
\begin{equation}
\Gamma \left(
\tilde{\nu}_{i} \to l^{+}_{j}l^{-}_{k}
\right) = \frac{1}{16 \pi}(\lambda_{ijk})^2m_{\tilde{\nu}_i},
\label{fcnc}
\end{equation}
where $m_{\tilde{\nu}_i}$ is the mass of sneutrinos and this
decay violate lepton number conservation. We have another
interesting LSP decay
\begin{eqnarray}
\tilde{\chi}^{0}_{1} \to \bar{l}_{i}l_{j}\nu_{k},
\end{eqnarray}
and the decay from the lighest neutralino, the Dark Matter
candidate at Minimal Supersymmetric Standard Model, produce
missing energy in its decay. This decay can be observed if
the appropriate $\lambda$ coupling satisfy the following
relation \cite{dress}
\begin{eqnarray}
| \lambda |>5 \times 10^{-7} \left(
\frac{m_{\tilde{l}}}{100 {\mbox GeV}} \right)^{2} \left(
\frac{100 {\mbox{GeV}}}{M_{LSP}}
\right)^{5/2},
\end{eqnarray}
where $m_{\tilde{l}}$ is the mass of charged slepton exchanged
and $M_{LSP}$ is the mass of LSP. However, if this sort of decays are not obaerved $\lambda$'s are smaller and the contribution to the neutrino masses become mainly from the interactions in Eq.~(\ref{yukawasusy1}).


\section{Conclusions}
\label{sec:con}

Here we have proposed a model for the  scotogenic mechanism with a global $B - L\times \mathbb{Z}_2$   symmetry. We guarantee that the scalar potential is bounded from below, after this we considered a inicial study on lepton flavor violation and dark matter, where the stability is guaranteed by the $\mathbb{Z}_2$   symmetry, in the first one we show that we do not have any restrictions and in the second we show that we are able to have good dark matter candidates. These models has several possibilities for dark matter but we will not consider this issue here. We have considered also the supersymmetric extension of the model and noted that in this case the neutralino contributions to the neutrino masses may be as important as the contributions existing in the non supersymmetric version.

Since the model is free of anomalies~\cite{Montero:2007cd,Ma:2014qra}, the $B-L$ symmetry could be gauged if more scalar fields are introduced, for instance two more active doublets, $S_{2,3}$ with $B-L$ charges $-8$ and $+10$, denoted $\phi$ and $\varphi$, respectively, in such a way that the quartic couplings $\lambda_9(S^\dagger_1D_1)(S^\dagger_2 D_1)$ and $\lambda_{10}(S^\dagger_1D_2)(S^\dagger_3 D_2)$ are possible (here $S_1=S$). If $B-L$ is gauged, terms as $\lambda_9$ and $\lambda_{10}$ can also arise from non-renormalizable interactions. For instance, by introducing two complex scalar singlet $\phi$ and $\varphi$ with $B-L=-8,10$, respectively, the dimension 5 interactions $(\lambda_9/\Lambda)(S^\dagger D_1)^2\phi$ and
$(\lambda_{10}/\Lambda)(S^\dagger D_2)^2\varphi$ induce those interactions, after the singlets get a non-zero VEV. These cases will be considered elsewhere.

\acknowledgments

ACBM thanks CAPES for financial support. JM thanks FAPESP for partial financial support at the beginning of this work and  IFT-UNESP for the kind hospitality where part of this work was done, also thanks C\'atedras CONACYT project 1753,  VP would like to thank CNPq for partial support and is also thankful for the support of FAPESP funding Grant No. 2014/19164-6, and MCR
thanks IFT-UNESP for the kind hospitality where part of this work was done.


\newpage

\end{document}